\documentclass[preprint]{revtex4-2}

\usepackage{graphicx}
\usepackage{dcolumn}
\usepackage{bm}
\usepackage{textcomp}

\begin{document}


\title{Demonstration of Autonomous Emittance Characterization at the Argonne Wakefield Accelerator}

\author{Ryan Roussel$^{1}$, Auralee Edelen$^{1}$, Dylan Kennedy$^{1}$, Seongyeol Kim$^2$, Eric Wisniewski$^2$ and John Power$^{2}$}
\email{rroussel@slac.stanford.edu;}

\address{%
$^{1}$ \quad SLAC National Accelerator Laboratory\\
$^{2}$ \quad Argonne National Laboratory}




\begin{abstract}
Transverse beam emittance plays a key role in the performance of high brightness accelerators. Characterizing beam emittance is often done using a quadrupole scan, which fits beam matrix elements to experimental measurements using first order optics. Despite its simplicity at face value, this procedure is difficult to automate due to practical limitations. Key issues that must be addressed include maintaining beam size measurement validity by keeping beams within the radius to diagnostic screens, ensuring that measurement fitting produces physically valid results, and accurately characterizing emittance uncertainty. We describe a demonstration of the Bayesian Exploration technique towards solving this problem at the Argonne Wakefield Accelerator, enabling a turnkey, autonomous quadrupole scan tool that can be used to quickly measure beam emittances at various locations in accelerators with limited operator input.
\end{abstract}

\maketitle

\section{Introduction}

Particle accelerators are complex instruments which require constant operator supervision and control to produce high-quality beams for use in a variety of scientific endeavours.
Often, this requires measuring beam attributes that have a high impact on accelerator applications, most notably, the transverse beam emittance.
A common method of measuring beam emittance is through the use of a tomographic manipulation of the beam distribution, commonly referred to as a quadrupole scan \cite{wiedemann_particle_2007}.

Despite the relative simplicity of the measurement technique, implementing it in practice faces a number of practical challenges.
Quadrupole focusing strengths must be chosen such that the beam remains within the confines of the diagnostic screen to guarantee that measurements of the beam size are accurate.
On the other hand, a wide range of focusing strengths must be used to sample multiple tomographic phase advances in order to accurately calculate the beam emittance.
Finally, fitting noisy experimental data to simplified, analytical models of beam transport can often result in non-physical results (imaginary emittances) and be subject to numerical errors due to catastrophic cancellation, making accurate predictions of the beam emittance, especially with calibrated uncertainty estimates, extremely challenging.
As a result of these numerous factors, emittance measurements using the quadrupole scan technique require substantial operator oversight to plan and validate.
This expends valuable resources, including beam and expert operator time, to perform routine diagnostic measurements.

In this work, we introduce and demonstrate a ``turn-key'' technique for robust, autonomous characterization of beam emittances with calibrated uncertainty estimates that requires little-to-no operator oversight.
Our method uses Bayesian active learning to autonomously choose quadrupole focusing strengths that maximize information gain about the beam size response.
Furthermore, our method tracks and builds a model of various observation constraints, particularly maximum beam size and location on the diagnostic screen, and incorporates this information into the measurement decision process.
We analyze the experimental data using physics-informed Bayesian inference to produce accurate predictions of the beam emittance with calibrated uncertainty estimates. 
Finally, we describe a demonstration of this technique at the Argonne Wakefield Accelerator.

\section{Materials and Methods}
\begin{figure}
   \centering
   \includegraphics[width=\columnwidth]{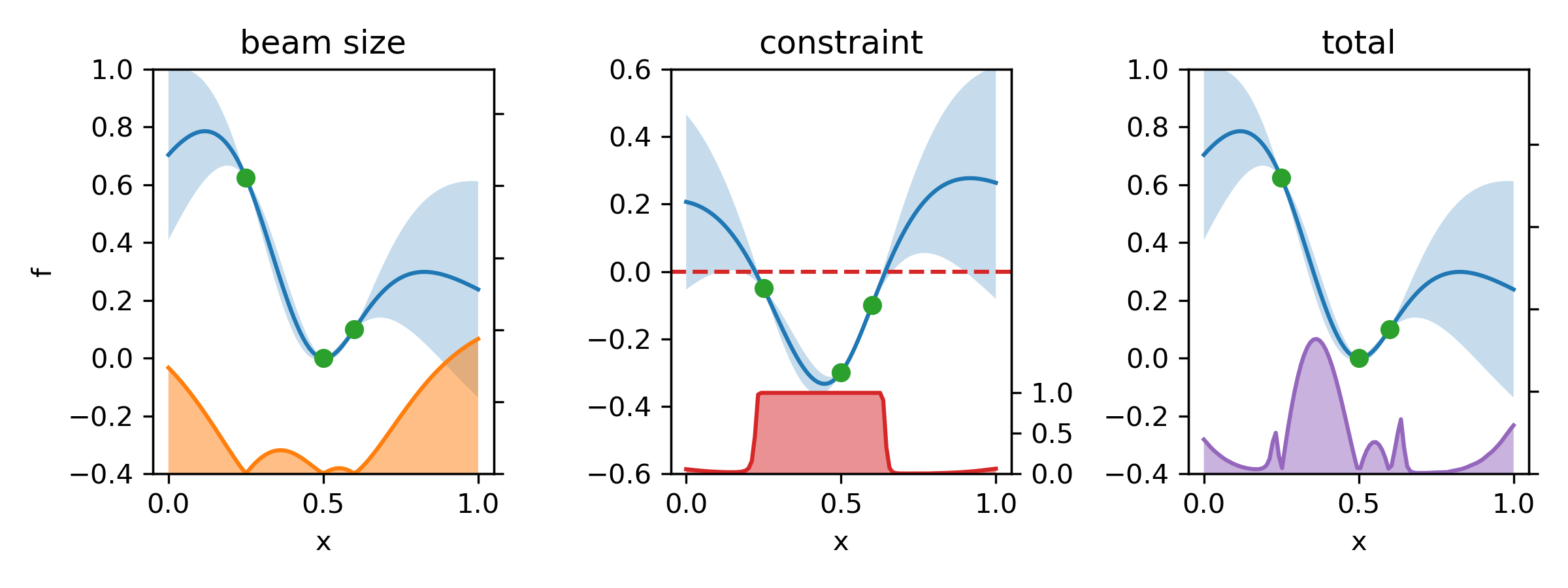}
   \caption{Illustration of the Bayesian exploration algorithm for characterizing unknown functions. Left: The uncertainty sampling acquisition function (orange) is equal to the model uncertainty (light blue) based on previous observations (green). This acquisition function is maximized at points furthest from previous measurements. Center: The constraint acquisition function term measures the probability that a constraint is satisfied (in this illustration when the constraining function value is less than zero). Right: The total acquisition function comprised of both uncertainty and constraint terms.}
   \label{fig:bayesian_exp}
\end{figure}

We combine a number of Bayesian techniques to create an autonomous emittance measurement algorithm known as Bayesian Exploration (BE) \cite{roussel_turn-key_2021}.
The foundation of our method is uncertainty sampling, illustrated in Figure \ref{fig:bayesian_exp} (left), which aims to choose measurements that provide the most information about the target function.
Bayesian exploration maintains an internal Gaussian process \cite{rasmussen_bayesian_2006} model of the target function and chooses measurements that maximize model uncertainty.
As a result for 1-D problems, this algorithm samples tuning configurations in a quasi-grid like fashion based on the configuration of initial measurements.
In higher dimensions, sampling becomes more sophisticated by incorporating the learned sensitivities of the target function towards individual tuning parameters. 
This approach intelligently enhances the sample density along axes that exhibit the strongest dependency, resulting in a more accurate representation of the underlying relationships with fewer measurements.

In addition to this intelligent sampling strategy, BE also considers observational constraints that need to be satisfied during characterization.
For quadrupole emittance scans, the primary constraint involves keeping the beam within a region of interest on the diagnostic screen to ensure valid beam size measurements.
Beams can extend outside a region of interest for several reasons, including under or over focusing of the beam and beam centroid deflection due to quadrupole misalignment's.
As a result, the range of quadrupole strengths that can be chosen for the emittance scan is limited by these factors to ensure valid measurements of the beam size.
Bayesian exploration handles these observational constraints by building independent models of these constraining functions and weighting the acquisition function by the probability that proposed measurements satisfy all of the constraints, as shown in Figure \ref{fig:bayesian_exp} (center, right).
Once models of the constraining functions reach a sufficient level of accuracy, BE will choose observations in a so-called ``valid'' sub-domain of the input space that are likely to satisfy observational constraints.
For example, in the toy case shown in Figure \ref{fig:bayesian_exp}, while the maximum uncertainty is at the edges of the input domain the constraint model predicts that these points have a low probability of satisfying the constraint, thus biasing sampling towards valid measurements closer to the domain interior.

\begin{figure}[b]
   \centering
   \includegraphics[width=0.35\columnwidth]{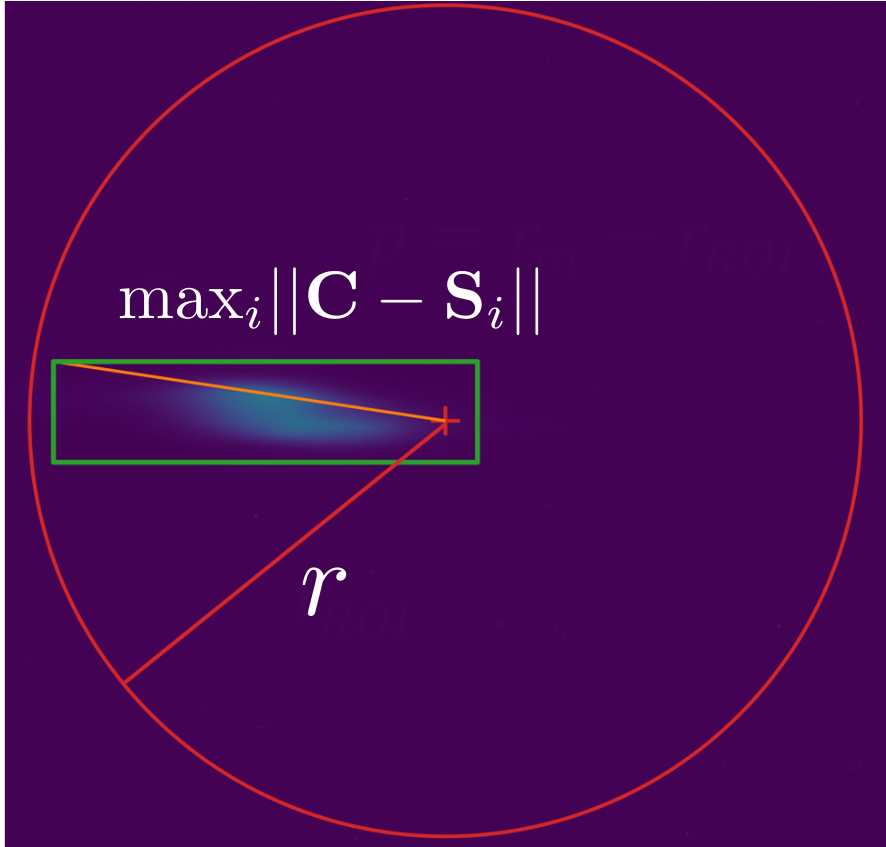}
   \caption{Diagram showing bounding box style image constraint.}
   \label{fig:image_constraint}
\end{figure}

We developed a specific constraining function in order to effectively reduce the frequency of invalid beam size measurements in the context of imaging diagnostics.
For BE to effectively predict where the valid sub-domain is in input space, it is critical that constraining functions are relatively smooth, such that functional correlations exist at nearby points in input space.
To satisfy this requirement, we use what we will refer to here as a ``bounding-box'' constraint, as shown in Figure \ref{fig:image_constraint}.
We specify a circular region of interest (ROI) in screen images with a center pixel coordinate $\mathbf{C}$ and a radius $r$ (also given in pixels).
After processing the raw screen image of a beam (using a Gaussian smoothing filter and a fixed minimum threshold) we calculate the weighted centroid and RMS size of the beam intensity inside the ROI in both the vertical and horizontal directions.
We then create a rectangular bounding box centered at the beam centroid with side lengths equal to 4 times the RMS beam sizes in each direction, which encapsulates most if not all of the beam intensity on the screen for observed beams.
The constraint function is then defined by the maximum distance between the ROI center and the bounding box corners, $c = \textbf{max}_i||\mathbf{C} - \mathbf{S}_i|| - r$, where $\mathbf{S}_i$ denotes pixel coordinates of each bounding box corner.
If the beam bounding box is inside the circular ROI then this constraining function is negative, conversely if it extends beyond the bounding box boundary then the constraining function value is positive.
If no beam is present on the screen, measured by specifying a minimum total intensity of image pixels, the constraining function value is set to a large value (in our case $c=1000$).

\subsection{Calculating emittances}
We calculate the beam emittance and corresponding uncertainty using samples drawn from a Gaussian Process model of the beam size squared.
The Gaussian process model is defined with a second order polynomial kernel such that samples drawn from the model are quadratic (equivalent to performing Bayesian regression using a second order polynomial function).
We then draw a large number of posterior samples from this model, and fit beam matrix elements to each sample independently, producing a corresponding distribution of possible emittance values. 
While the quadratic kernel ensures that posterior samples from the Gaussian process model are quadratic with respect to the inputs, there is no guarantee that sampled parabolas are physically valid. 
Therefore, we reject samples that predict negative beam sizes and imaginary emittances.
The remaining samples then form the probability distribution of physically valid emittance values.

\subsection{Experimental Demonstration}
We conducted an experimental demonstration of automatic emittance measurements at the Argonne Wakefield Accelerator (AWA).
Our study attempted to characterize the beam emittance of beams exiting the accelerating section of the AWA beamline using a single quadrupole magnet (effective length 0.12 m) and a YAG diagnostic screen located 1.065 m downstream.
First, the beam was centered on the screen and manipulated by upstream quadrupoles to fit within the ROI.
Then, we use the python library Xopt \cite{roussel_xopt_2023} to sample 4 chosen points to create an initial data set.
Xopt was then used to perform constrained Bayesian Exploration as described in the previous sections with a Gaussian process using a default Matern kernel.
After a fixed number of iterations, in this case 20, the algorithm was terminated and data was used to calculate a distribution of possible horizontal emittances.

\section{Results}
\begin{figure}[h]
   \centering
   \includegraphics[width=\columnwidth]{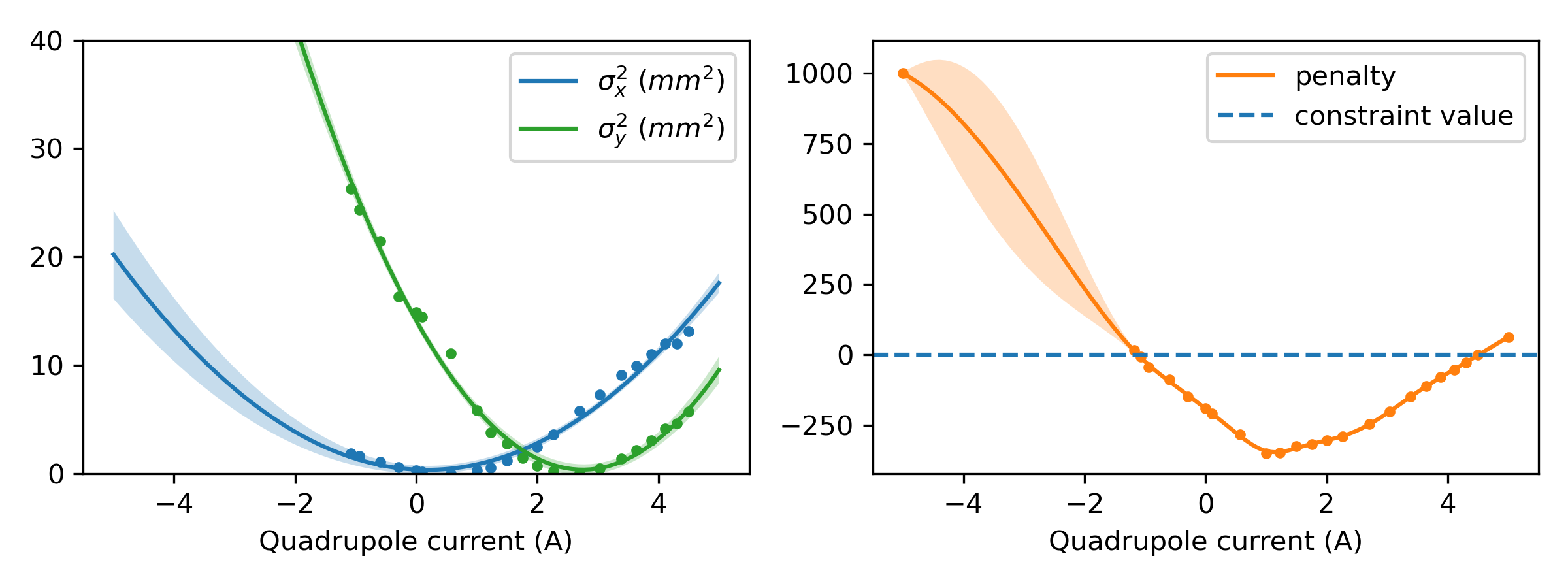}
   \caption{Left: Beam sizes squared plotted for both horizontal and vertical directions as a function of quadrupole current, along with Bayesian model fits using second-order polynomial fits. Right: Constraining function measurements and model predictions. Dashed line denotes maximum allowed value of the penalty function which satisfies the constraint.}
   \label{fig:auto_quad_scan}
\end{figure}

\begin{figure}
   \centering
   \includegraphics[width=\columnwidth]{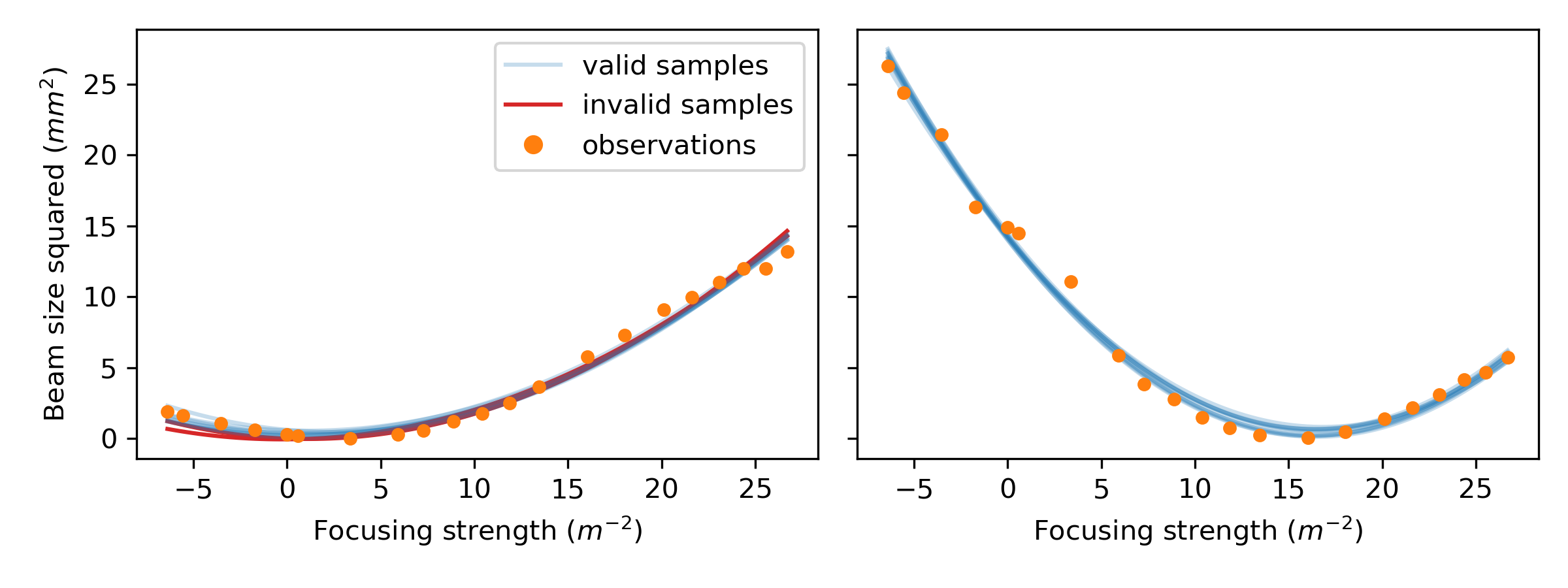}
   \caption{Plots showing samples drawn from the Gaussian process model used to determine the beam emittance for the horizontal (left) and vertical (right) axes. Samples that correspond to imaginary emittances are denoted as ``invalid" and are rejected when calculating the distribution of predicted emittances.}
   \label{fig:scan_samples}
\end{figure}

\begin{figure}
   \centering
   \includegraphics[width=0.65\columnwidth]{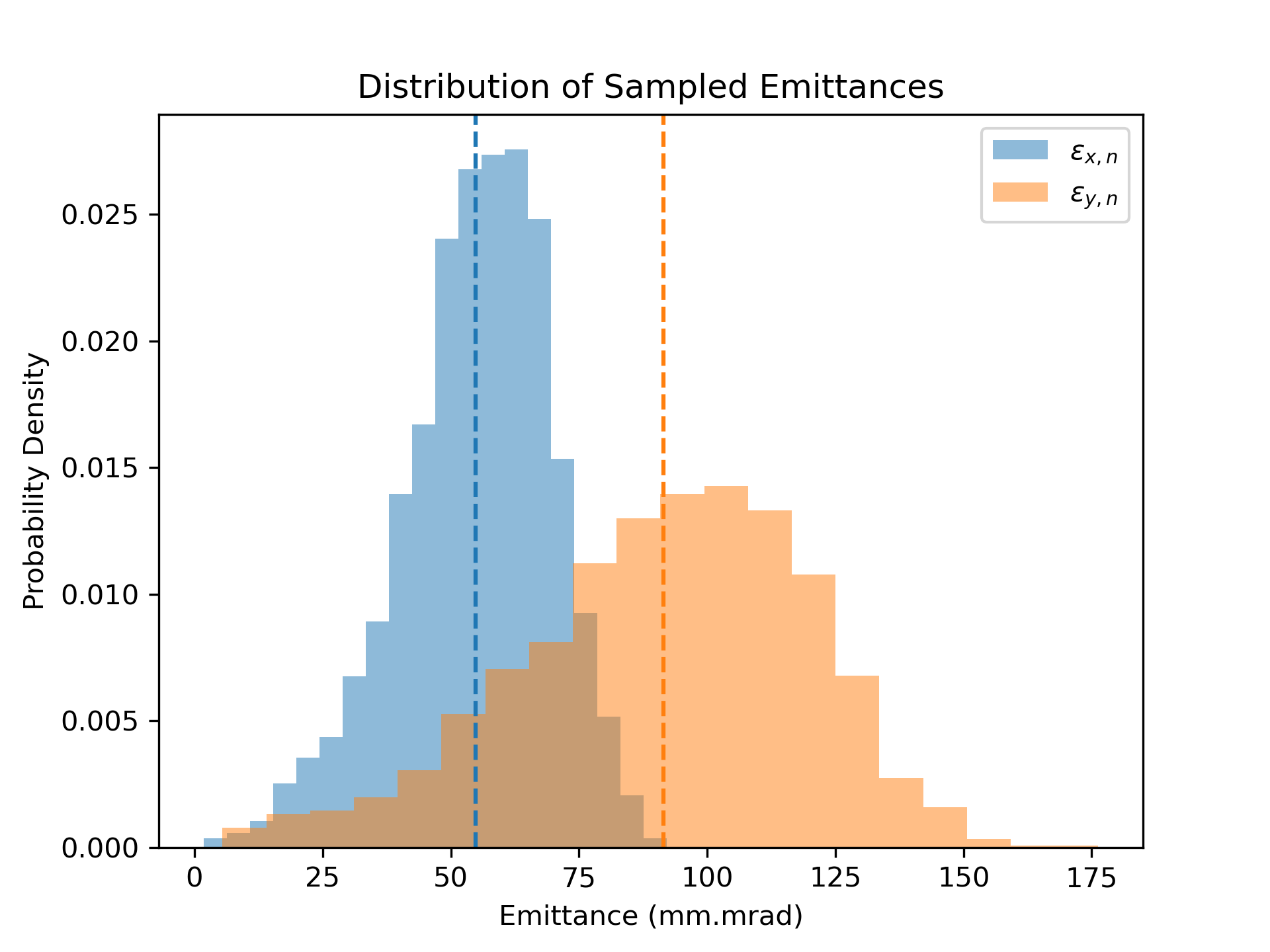}
   \caption{Distribution of horizontal and vertical emittances calculated using samples drawn from the predictive model.}
   \label{fig:emittance_samples}
\end{figure}

Results from the experimental demonstration are shown in Figures~ \ref{fig:auto_quad_scan}-\ref{fig:emittance_samples}.
In Figure~\ref{fig:auto_quad_scan}(left) we observe that Bayesian exploration distributed beam size measurements evenly throughout the valid input space of quadrupole strengths.
Figure \ref{fig:auto_quad_scan}(right) shows that the constraining function is learned during the exploration process, resulting in only 3 measurements that violated the  constraint.

Figure \ref{fig:scan_samples} shows samples drawn from the GP posterior model.
An emittance value is calculated for each sample using a second order polynomial fit to calculate elements of the beam matrix.
Samples that predict an imaginary beam emittance (approx. 10\% for the data sets shown here) are considered ``invalid" and are rejected.
Figure \ref{fig:emittance_samples} shows predictions of the beam emittance from valid samples drawn from the predictive beamsize model.
We observe a horizontal emittance of $\varepsilon_{x,n} = 54.77 \pm 14.63$ mm.mrad and a vertical emittance of $\varepsilon_{y,n} = 91.42 \pm 28.01$ mm.mrad.
These predictions are reasonable given an un-optimized drive beamline with 1 nC bunch charge at the AWA facility.
Despite relatively small uncertainty in the predictive beamsize model, there is still significant uncertainty in the beam emittance.
It is possible that these large uncertainties are a result of a combination of factors, including noisy measurements, limited phase space advances due to constraining functions, and catastrophic cancellation effects.

\section{Discussion}
Our results show that this algorithm is successful in automating the quadrupole scan process given arbitrary upstream beamline parameters, thus reducing the burden on accelerator operators when emittance measurements are needed.
The algorithm can select quadrupole strengths to rotate the beam in phase space while adhering to practical constraints.

This method can be further improved through several means. 
First, beam size measurements at every shot can be used in creating the predictive model, as opposed to using averaged measurements, which would improve the accuracy of emittance uncertainty estimates due to jitter.
Second, the speed of decision-making in the algorithm could be increased by using a mesh numerical optimizer of the acquisition function, since the decision space is only one-dimensional.
Finally, instead of using the beam images to calculate RMS beam sizes for fitting a polynomial model, the entire image can be used to accurately reconstruct the transverse phase space distribution, as is done in \cite{PhysRevLett.130.145001}.

\vspace{6pt} 



\section{Acknowledgments}
This work was funded by the U.S. Department of Energy, Office of Science, Office of Basic Energy Sciences under Contract No. DE-AC02-76SF00515.

Conceptualization, Ryan Roussel and Auralee Edelen; Data curation, Ryan Roussel; Formal analysis, Ryan Roussel; Funding acquisition, Auralee Edelen; Investigation, Ryan Roussel, Seongyeol Kim and Eric Wisniewski; Methodology, Ryan Roussel, Dylan Kennedy and Auralee Edelen; Software, Ryan Roussel and Dylan Kennedy; Supervision, Auralee Edelen and John Power; Validation, Ryan Roussel; Visualization, Ryan Roussel; Writing – original draft, Ryan Roussel, Dylan Kennedy and Auralee Edelen; Writing – review \& editing, Ryan Roussel, Dylan Kennedy and Auralee Edelen.


\bibliography{auto_quad_scan}

\end{document}